\documentclass[aps,prl,twocolumn,groupedaddress]{revtex4}

\def\gap{\hbox{${_{\displaystyle>}\atop^{\displaystyle\sim}}$}}

\def \be{\begin{equation}}
\def \ee{\end{equation}}
\def \beq{\begin{eqnarray}}
\def \eeq{\end{eqnarray}}

\newcommand{\rxa}{\hbox{RX\,J0420.0$-$5022}}
\newcommand{\rxb}{\hbox{RX\,J0720.4$-$3125}}
\newcommand{\rxc}{\hbox{RX\,J0806.4$-$4123}}

\newcommand{\rxe}{\hbox{RX\,J1605.3+3249}}
\newcommand{\rxf}{\hbox{RX\,J1856.4$-$3754}}

\newcommand{\rbsd}{\hbox{RBS1223}}
\newcommand{\rbsg}{\hbox{RBS1774}}
\newcommand{\axpa}{\hbox{CXOU\,J0100-7211}}
\newcommand{\axpb}{\hbox{4U\,0142+61}}
\newcommand{\axpc}{\hbox{1E\,1048.1-5937}}
\newcommand{\axpd}{\hbox{1RXS\,J1708-4009}}
\newcommand{\axpe}{\hbox{XTE\,J1810-197}}
\newcommand{\axpf}{\hbox{1E\,1841-045}}

\newcommand{\axph}{\hbox{1E\,2259+586}}
\newcommand{\sgra}{\hbox{SGR\,0526-66}}

\newcommand{\sgrd}{\hbox{SGR\,1900+14}}

\usepackage{graphicx}

\begin{document}


\title{Evidence for Heating of Neutron Stars by Magnetic Field Decay} 


\author{Jos\'e A.~Pons}
\altaffiliation{Departament de F\'{\i}sica Aplicada, Universitat d'Alacant,
           Ap. Correus 99, 03080 Alacant, Spain}
\author{Bennett Link}
\altaffiliation{Montana State University, Department of Physics, Bozeman MT
59717}
\author{Juan A.~Miralles}
\altaffiliation{Departament de F\'{\i}sica Aplicada, Universitat d'Alacant,
           Ap. Correus 99, 03080 Alacant, Spain}
\author{Ulrich Geppert}
\altaffiliation{Departament de F\'{\i}sica Aplicada, Universitat d'Alacant,
           Ap. Correus 99, 03080 Alacant, Spain}

\affiliation{}


\date{\today}

\begin{abstract}
We show the existence of a strong trend between neutron star (NS)

surface temperature and the dipolar component of the magnetic field
extending through three orders of field magnitude, a range that
includes magnetars, radio-quiet isolated neutron stars, and 
many ordinary radio pulsars. We
suggest that this trend can be explained by the decay of currents in
the crust over a time scale of $\sim 10^6$ yr.  We estimate the
minimum temperature that a NS with a given magnetic field can reach in
this interpretation.
\bigskip\bigskip
\end{abstract}


\bigskip

\maketitle


A question of fundamental importance in subatomic physics concerns the
ground state and emissivity of dense matter in beta equilibrium. In
this connection, the manner in which a neutron star (NS) cools after
its birth has been an active area of research since the discovery of
NSs as pulsars four decades ago.  Cooling occurs through neutrino
emission for the first $\sim 10^5$ yr of its life, and later by
surface thermal emission (see, e.g., \cite{pr06} for a review).  As
the star loses its residual heat, any internal heat sources would
affect, and possibly control, the star's thermal evolution. One
important heat source could be decay of the star's magnetic field if
it occurs over a sufficiently rapid time scale. The field could decay
directly as a consequence of the non-zero resistivity of the matter
(Ohmic decay) or ambipolar diffusion, or indirectly as a consequence
of Hall drift that produces a cascade of the field to high wave number
components that decay rapidly through Ohmic decay
\cite{gr92,rg02}. The possibility of field decay has motivated
extensive work to assess how the thermal evolution would be affected 
(see, e.g., \cite{haensel_etal90,MUK98} and
references therein). Ohmic decay is expected to proceed most rapidly
in the crust, where the conductivity is determined primarily by
electrons colliding with phonons and impurities \cite{fi76}. With
considerable uncertainties in conductivities and transport
coefficients, clear conclusions as to the effects of Ohmic decay have
not been reached, though it appears likely that it could play
some role in the thermal evolution of NSs.
Magnetic field evolution is,
however, expected to play a key role in the evolution of {\em
magnetars}, NSs with fields $\gap 10^{14}$ G. Magnetars are
remarkable in the sense that their magnetic energy exceeds their
rotational energy, in contrast to the lower-field {\em
rotation-powered} pulsars. In magnetars, dissipative field evolution 
is expected to occur, and the magnetic energy available is so great that
substantial energy can be dissipated, contributing to the star's
heat budget. 

In this Letter, we present observational evidence for a strong
correlation between stellar magnetic field and surface temperature. We
suggest that this trend can be simply explained by energy dissipation
from field decay in the crust over a time scale of $\sim 10^6$ yr. We
argue that NSs with fields $\gap 10^{13}$ G begin to have their
thermal evolution controlled by field decay about when they enter the
photon cooling era at $\sim 10^5$ yr, and that magnetars, which have
more magnetic energy available, are dominated by field decay even
earlier. Our conclusions are essentially independent of the
uncertainties concerning stellar structure and the state of matter
above nuclear saturation.

\begin{table}
\caption{Properties of NSs with reported
thermal emission.  $B_d$ is estimated by assuming spin-down from dipole 
radiation, $B_d = 3.2 \times 10^{19} \sqrt{P \dot{P}}$ G, where $P$ is the
spin period and $\dot{P}$ is its time derivative. Except for
\rxf$\,$ (see footnote), $B_{o}$ was estimated 
assuming observed x-ray absorption features are proton cyclotron 
lines. Ages are spin-down ages ($P/2\dot{P}$) except for \rxf.
}
\begin{tabular}{lcccc}
\hline
\hline
\noalign{\smallskip}
 Source & kT & B$_{d}$ (B$_{o}$) & Age  & Ref. \\
        & (keV) & (10$^{13}$ G) & (kyr) & \\
\noalign{\smallskip}\hline\noalign{\smallskip}
Magnetars  & & \\
\noalign{\smallskip}\hline\noalign{\smallskip}
\sgra & 0.53  &  74 & 1.9 & \cite{WT06} \\
\sgrd & 0.43  &  57 & 1.3 & \cite{WT06} \\
\axpa & 0.38  &  39 & 6.8 & \cite{MGa05} \\
\axpb & 0.46  &  13 & 70  & \cite{WT06} \\
\axpc & 0.63  &  39 & 1-8 & \cite{WT06} \\
\axpd & 0.44  &  47 & 9.0 & \cite{WT06} \\
\axpe & 0.67  &  29 & 5.7 & \cite{WT06} \\
\axpf & 0.44  &  71 & 4.5 & \cite{WT06} \\
\axph & 0.41  &  6  & 220 & \cite{WT06} \\
\noalign{\smallskip}\hline\noalign{\smallskip}
 $P>3$ s & & \\
\noalign{\smallskip}\hline\noalign{\smallskip}
\rxa   & 0.044 & $<18$ (6.6) &  & \cite{haberl} \\
\rxb   & 0.090 & 2.4   (5.6) & 1900 & \cite{haberl} \\
\rxc   & 0.096 & $<14$ (6.1) & & \cite{haberl} \\
\rbsd  & 0.086 & 3.4   (4.6) & 1461 & \cite{haberl} \\
\rxe   & 0.096 &    (8.0) & & \cite{haberl} \\
\rxf   & 0.062 &     (1)
\footnote{In the case of \rxf$\,$ the magnetic field 
was estimated from the spin-down luminosity required
to power its $H_\alpha$ emission nebula \cite{Tru04}.}
  & (500) &\cite{haberl,Tru04} \\
\rbsg  & 0.102 & $<24$  (15) & & \cite{haberl} \\
CXOU J1819-1458 & 0.120  & 5.0 & 117 &\cite{Rey06} \\
PSR J1718-3718  & 0.145  & 7.4 & 34 &\cite{KM05} \\
PSR B2334+61    & 0.056  & 1.0 & 41 & \cite{McG06} \\
\noalign{\smallskip}\hline\noalign{\smallskip}
 $P<0.5$ s & & \\
\noalign{\smallskip}\hline\noalign{\smallskip}
Geminga         & 0.03-0.04      & 0.16  & 340 & \cite{Kar05,DeL05} \\
PSR B1055-52    & $\approx$0.06  & 0.11  & 530 & \cite{DeL05} \\
PSR B0656+14    & 0.059-0.12     & 0.467 & 110 & \cite{DeL05} \\
PSR J1119-6127  & 0.207          & 4.1   & 1.6 & \cite{Gon05} \\
Vela            & 0.056-0.061    & 0.34  & 11 & \cite{Vela} \\
PSR B1706-44    & 0.04-0.07      & 0.3   & 17 & \cite{McG04} \\
PSR J0205+6449  & $<$ 0.094      & 0.36  & 5 & \cite{Sla02} \\
Crab            & $<$ 0.17       & 0.38  & 1.2 & \cite{Wei04} \\
\noalign{\smallskip}\hline\noalign{\smallskip}
\end{tabular}
\label{tab1}
\end{table}

To evaluate the extent to which the magnetic field of a star
determines its temperature, we show in Fig. 1 the effective surface
temperature $T_{\rm eff}$ vs. the dipole component of the magnetic
field $B_d$ estimated for 27 NSs (Tab. 1). We note a striking trend of
$T_{\rm eff}$ with $B_d$ well approximated by $T_{\rm eff}\propto
B_d^{1/2}$. This trend holds over three orders of
magnitude in $B_d$, encompassing much of the observed range of 
magnetic fields.  Fig. 1 suggests that the thermal evolution
of NSs with $B\gap 10^{13}$ is largely determined by the
strength of the magnetic field.

The spectra of some stars are well-described by a simple blackbody
(BB) associated with thermal surface emission. In many stars, however,
the spectrum comprises both a magnetospheric component and one or two
BB components. Two-component BB spectra are indicative of temperature
anisotropy over the stellar surface, presumably smooth, but modeled as
being relatively cold with small hot spots around the magnetic poles.
For pure BB stars, we took $T_{\rm eff}$ to be the measured
temperature of an unknown area $A_{\rm eff}$ of the stellar
surface. For stars with spectra that include two BB components, we
used the temperature of the component which dominates the spectrum
obtained from the references cited in Tab. 1.  Some reported
temperatures are not BB temperatures, but were obtained with specific
atmospheric models (mainly H atmospheres); atmospheric compositional
uncertainties introduce an uncertainty of a factor of $\sim 2$ in
$T_{\rm eff}$, which is unimportant for our purposes. Fig. 1 contains
a point for every star from which thermal emission has been observed
with reasonable confidence. These include: ordinary radio pulsars;
isolated NSs which show no radio emission but thermal X-ray emission,
and magnetars.

\begin{figure}
\resizebox{\hsize}{!}{\includegraphics{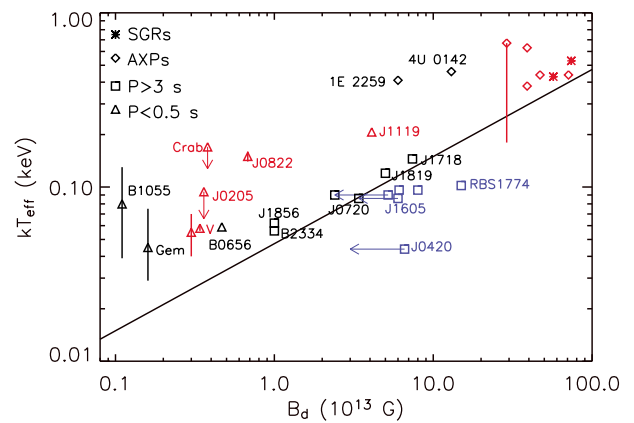}}
\caption{$T_{\rm eff}$ vs. $B_d$ of isolated NSs. 
Represented with different symbols are SGRs (stars), AXPs (diamonds), 
slowly-rotating ($P>3$ s) NSs (squares) and rapidly-rotating ($P<0.5$ s) NSs
(triangles). Red symbols correspond to young ($<10^4$ yr) NSs.
Symbols with arrows indicate upper limits. 
The blue squares are isolated NSs for which the magnetic field
was estimated from the association of a spectral feature with a proton
cyclotron resonance. We show how two of these (\rxb\ and RBS1223) move to
the left if their fields inferred from cyclotron lines are replaced by $B_d$.
The solid line is the is an illustration of heating balanced by
cooling, for $b=100$ (see eq. \ref{hbl}).
} 
\label{pulsar}
\end{figure}

We consider it highly unlikely that the observed distribution of
$T_{\rm eff}$ along a narrow diagonal band is a selection
effect. If there were no relationship between $T_{\rm eff}$ and $B_d$,
we would expect to see many examples of stars in the band between
0.03-0.2 keV with low fields ($\lesssim 10^{12}$ G) since we already
see such hot stars at high fields ($\gtrsim 10^{13}$ G), and stars
with fields of $\sim 10^{12}$ G are $\sim 100$ times more numerous
than those with fields of $\sim 10^{13}$ G.  The dearth of hot stars
at fields below $\sim 10^{12}$ G, therefore, is almost certainly not due to a
selection effect. While it is true that very high-field stars are rare
and therefore more distant on average, giving a preference to seeing
objects with high $T_{\rm eff}$, continuing surveys of higher
sensitivity have failed to reveal sources cooler than shown in Fig. 1
above $10^{13}$ G after years of observation.  Taken together, these
facts strongly suggest that the trend we are seeing is real, though
population simulations might be able to provide a definite answer. The
natural interpretation of this diagram is that stars with fields of
$\sim 10^{12}$ G cool much more rapidly than stars with fields of
$\sim 10^{13}$ G and higher. It is generally believed that magnetars
are kept hot by decay of their strong magnetic fields. We propose that
the same is happening in NSs with fields down to $\sim 10^{13}$ G.

Aside from the interpretation of bursts in magnetars as representing
large-scale field evolution and decay, there is no convincing
observational evidence for magnetic field decay in the NS population
as a whole.  Statistical studies of the entire NS population have
generally found that continuous exponential decay of the dipole
component of NS magnetic fields, if it occurs, cannot happen over time
scales shorter than $\sim 10^8$ yr ({\sl e.g.},
\cite{PS1}, but see \cite{gonthier}). 
We suggest that the general trend of Fig. 1 can be explained by the
decay of {\em crust currents} in stars with $B_d\gap 10^{13}$; these
stars constitute only $\sim 5$\% of the stellar population, so there
is no obvious conflict with the conclusions cited above against field
decay in the stellar population as a whole. Moreover, we do not claim
that $B_d$ decays indefinitely, the hypothesis those studies
considered.

If heating by decay of crust currents is relevant in the more
strongly-magnetized stars, a NS of some initial
magnetic field will initially cool through neutrino emission, but
eventually crustal field decay will dissipate enough energy to
contribute significantly to the star's photon emission. When this
happens depends on the strength of the initial field; the stronger the
field, the earlier its decay begins to control the surface emission.
Eventually, dissipation of the field will nearly balance loss to
surface thermal emission, and the thermal evolution will be
subsequently determined by this balance.  For illustration, we suppose
that the hot spot of area $A_{\rm eff}$ and temperature $T_{\rm eff}$
is kept hot by the dissipation of magnetic energy in a volume $A_{\rm
eff}\Delta R$ directly below it, where $\Delta R\simeq 1$ km is the
crust thickness.  The near balance between heating and cooling is
expressed by
\begin{equation}
-A_{\rm eff} \Delta R \frac{dE_m}{dt}=A_{\rm eff} \sigma T_{\rm
eff}^4,
\label{balance}
\end{equation}
where $E_m=B^2/8\pi$ is the magnetic energy density in the crust, $B$
is the field strength there and $\sigma$ is the Stefan-Boltzmann
constant. The unknown $A_{\rm eff}$ does not determine equilibrium in
this simple model. We expect that the basic energy scale of the
magnetic energy available in the crust for dissipation is set by the
dipole field, and we parameterize the crust field strength as
$B^2=bB_d^2$; $b$ is the ratio of magnetic energy density due to
currents in the crust to the dipole energy density. The crust field,
which presumably includes multipole and toroidal contributions,
is not directly observable, but modeling of the thermal spectra of
strongly magnetized NSs that show {\em only} thermal emission
\cite{paper1,GKP04} indicate $b\simeq 100$. 
These objects show a time-dependent flux, due to spin modulation of
emission from a hot spot, and an optical excess interpreted as the
tail of a much softer thermal emission that gives a negligible
contribution to the X-ray emission
\cite{Pons02}. This interpretation of the X-ray and optical data
implies the existence of a large degree of anisotropy in the surface
temperature, due to a magnetic field in the crust
that is large compared to $B_d$ and has significant toroidal components.

In heating-cooling equilibrium, the cooling history of the star will
be simply coupled to the decay of the magnetic field. Different
processes, such as Ohmic decay and Hall drift, can contribute to field
decay in crust. For purposes of illustration, we assume simple
exponential decay of the magnetic field over a time scale $\tau_D$,
\beq
\frac{d \mathbf B}{dt} = -\frac{\mathbf B}{\tau_D}
\label{model}
\eeq
which is equivalent to assuming that the magnetic energy density $E_m$
decays at a rate proportional to $E_m$. 
Combining eqs. [\ref{balance}] and [\ref{model}]:
\begin{equation}
\Delta R\, b\, B_d^2 = 4\pi\tau_D \sigma T_{\rm eff}^4.
\label{hbl}
\end{equation}
This simple model accounts for the trend $T_{\rm eff}\propto
B_d^{1/2}$ shown in Fig. 1; it gives a {\em heating balance line}
(HBL), along which older NSs should cluster. The location of the HBL
on Fig. \ref{pulsar} is determined by the product $\tau_D^{-1}b$. Each
star will have it's own HBL to the extent that $b$ varies among
different stars.  In Fig. 1 we show an example of one possible HBL
that approximately follows the data, corresponding to $\tau_D \simeq 5
\times 10^3 b$ yr.  For $b=100$ for example, we estimate $\tau_D\simeq 10^6$ yr
as the characteristic decay time. This time scale is comparable to the
Ohmic decay time estimated for an impure crust \cite{jones}. 

We now discuss how a NS reaches its HBL in this picture.  A NS will
begin its life high on Fig. 1 with some $B_d$.  As it cools it moves
vertically downward, until decay of its field causes the trajectory to
bend to the left. The star eventually reaches its HBL, and then
continues moving down it. This model predicts that no object will be
found below its HBL.  Well above the line, we should see only young
hot NSs following their respective cooling trajectories which are not
yet affected by heating from field decay.  Cooling simulations without
heating predict that the principal energy loss changes from
neutrinos to surface photon emission at an age of $\sim 10^5$ yr,
independent of the birth temperature (e.g., \cite{pr06}). Most stars
will not reach their HBLs until about this age, though very high-field
objects can reach their HBLs earlier as they have more magnetic
energy to dissipate. We have therefore plotted with red symbols those
objects with ages under $10^4$ yr for reference.  It is remarkable
that slowly rotating NSs as well as most magnetars all fall close to
the representative HBL.  Rapidly rotating NSs (such as PSR B1055) were
probably born with initially lower fields which implies a less
efficient spin down. They are still moving vertically in this diagram
because, due to their weaker field, heating from field decay is only
relevant at later times.  According to this picture, some old NSs {\em
could} be former magnetars, whose magnetic fields have decayed by a
factor of $\sim 10$. This evolutionary path was proposed for \rxb\
\cite{HK98}, but we suggest that it is more general and applies to
many other objects.


In our simple energy balance argument, we ignored the fact that some of the
dissipated energy will flow into the core and be lost to neutrino
emission. Kaminker et al \cite{Kam06}, for example, find that
continuous heat deposition in 1-d simulations without a magnetic field
is largely lost to neutrinos if the energy is deposited at densities
above neutron drip (the beginning of the inner crust). The strong
crustal fields we are proposing, however, will greatly suppress
heat transport into the core, while allowing efficient transport along
the field lines, which go almost directly to the surface. This effect
will be investigated further in future work. 

Some of the stars do fall slightly below our representative HBL, but
this is not surprising since $b$ should vary among stars and we show
here only one example. There are also uncertainties in $T_{\rm eff}$
and $B_d$. Some of the objects in Fig. 1, the blue squares, have
magnetic fields determined under the (not generally accepted)
assumption that the absorption lines in their spectra are proton
cyclotron lines, which
should give estimates of the field larger than the dipolar
component. For two cases in which $B_d$ is also known from $P{\dot P}$
(\rxb\ and RBS1223), use of $B_d$ brings these objects onto our
example HBL (see Fig. 1).  We also note that two magnetars, \axph\ and
\axpb, while following the general trend of Fig. 1, lie above our
representative HBL. These objects show frequent burst activity and
complex evolution of their light curves. If these objects are
releasing magnetic energy episodically there
would be additional heating occurring, increasing $T_{\rm eff}$ above
what we would expect in our scenario of gradual field decay. 

We have argued that the strong dependence of $T_{\rm eff}$ on $B_d$
for stars with $B\gap 10^{13}$ G (Fig. 1) indicates that the thermal
evolution is almost completely controlled by the amount of magnetic
energy the star has stored in its crust by the time the star has
reached an age of $\sim 10^5$ yr (earlier, for magnetars).  This
conclusion is insensitive to uncertainties about the state of the
stellar core, its structure and the rates of neutrino processes that
take place there. The specific heat and thermal conductivity through
the star are also unimportant, provided that field decay does occur,
and that the energy liberated emerges primarily at the stellar
surface. The data are consistent with the decay of crustal fields
about an order of magnitude stronger than the dipole component, over
$\sim 10^6$ yr in all stars. It appears that the effects of strong
crust fields, heat generation from their decay, and modified heat
transport in the crust should all be considered towards obtaining a
more complete understanding of NS cooling. The evidence for crustal
field decay presented here also has implications for estimates of the
ages of pulsars older than $\tau_D$; the standard spin-down age,
$P/2\dot{P}$, then significantly overestimates the star's true age.

\begin{acknowledgments}
We thank A. Cumming for interesting discussions.
This work has been supported by the Spanish MEC grant AYA 2004-08067-C03-02.
JAP is supported by a {\it Ram\'on y Cajal} contract.
B. L. acknowledges support from U. S. NSF grant AST-0406832. 
U.G. acknowledges support from the Spanish MEC program SAB-2005-0122.
\end{acknowledgments}

\end{document}